\documentclass[aps,pra,showpacs,amsmath,preprint,amssymb]{revtex4-1}

\usepackage[T1]{fontenc}
\usepackage[latin9]{inputenc}
\usepackage{color}
\usepackage{graphicx}
\usepackage{url}
\usepackage[normalem]{ulem}
\usepackage{amsthm}
\usepackage{amsmath}

\def\be{\begin{equation}}
\def\ee{\end{equation}}
\def\bea{\begin{eqnarray}}
\def\eea{\end{eqnarray}}
\def\bi{\begin{itemize}}
\def\ei{\end{itemize}}
\def\bin{\begin{enumerate}}
\def\ein{\end{enumerate}}

\def\ra{\rangle}

\newcommand{\vect}[1]{\mathbf{#1}}

\begin{document}
\title{Artificial magnetic field induced by an evanescent wave}

\author{Ma\l{}gorzata Mochol}
\affiliation{
Instytut Fizyki imienia Mariana Smoluchowskiego and
Mark Kac Complex Systems Research Center, 
Uniwersytet Jagiello\'nski, ulica prof. Stanis\l{}awa \L{}ojasiewicza 11, PL-30-348 Krak\'ow, Poland}

\author{Krzysztof Sacha}
\affiliation{
Instytut Fizyki imienia Mariana Smoluchowskiego and
Mark Kac Complex Systems Research Center, 
Uniwersytet Jagiello\'nski, ulica prof. Stanis\l{}awa \L{}ojasiewicza 11, PL-30-348 Krak\'ow, Poland}

\date{\today}

\begin{abstract}
Cold atomic gases are perfect laboratories for realization of quantum simulators. In order to simulate solid state systems in the presence of magnetic fields special effort has to be made because atoms are charge neutral. There are different methods for realization of artificial magnetic fields, that is the creation of specific conditions so that the motion of neutral particles mimics the dynamics of charged particles in an effective magnetic field. Here, we consider adiabatic motion of atoms in the presence of an evanescent wave. Theoretical description of the adiabatic motion involves artificial vector and scalar potentials related to the Berry phases. Due to the large gradient of the evanescent field amplitude, the potentials can be strong enough to induce measurable effects in cold atomic gases. We show that the resulting artificial magnetic field is able to induce vortices in a Bose-Einstein condensate trapped close to a surface of a prism where the evanescent wave is created. We also analyze motion of 
an atomic cloud released from a magneto-optical trap that falls down on the surface of the prism. The artificial magnetic field is able to reflect falling atoms that can be observed experimentally.
\end{abstract}


\maketitle

\section*{Introduction}
Cold atomic gases are flexible laboratories with a great potential to investigate a variety of problems from many fields of physics \cite{lewenstein2007,bloch2008}. Trapping potentials for atoms and mutual atom interactions can be controlled and engineered nearly at will. Mixtures of different atomic species of the Fermi and Bose statistics can be prepared and investigated experimentally. 

Atoms are charge neutral and seem not suitable to simulate orbital magnetism. However, there are also methods that allow for a generation of artificial gauge fields, i.e. the creation of specific conditions such that the motion of neutral particles mimics the dynamics of charged particles in an effective magnetic field, see \cite{galitski13,dalibard2011,juzeliunas13} and references therein. There are also proposals to simulate the spin-orbit coupling with the help of non-Abelian artificial gauge fields \cite{juzeliunas13,lin2011,wang2012,cheuk2012,zhang2012,fu2013,zhang2013,qu2013,leblanc2013,kosior2014}. Ultra-cold atoms in synthetic gauge fields open possibilities for deep understanding of fundamental problems like high-temperature superconductivity or strongly interacting counterparts of topological insulators \cite{hasan2010}.  

In the present article we focus on a method of the synthetic magnetic field creation which involves adiabatic motion of atoms in a laser radiation whose intensity changes in space \cite{juzel2008,dalibard2011}. We assume that atoms are placed close to a dielectric surface and experience an evanescent field that penetrates the region of the atoms location. Large gradient of the generalized Rabi frequency is responsible for the presence of geometrical vector and scalar potentials that are related to the Berry phases \cite{berry1984}. These potentials generate an artificial magnetic field for the charge neutral particles which is able to induce vortices in an atomic cloud cooled down to the quantum degeneracy.    

Mirrors are basic elements of optical devices. Similarly atomic mirrors that can reflect beams of atoms are of great experimental interest. They can base on the interaction between magnetic moments of atoms with magnetic fields \cite{magn_mirr} or take advantage of a spatially dependent light-shift of atomic energy levels. An evanescent field is able to create a strong optical dipole potential for atoms and is a convenient tool to build atomic mirrors \cite{westbrook98,ev_trap,gillen2009,cornelussen2002,fiutowski2013}. Evanescent fields can be created by shining laser radiation on a dielectric surface but they are also connected with surface plasmons that are collective oscillations of free eletrons at a metal surface \cite{plasmon}. Plasmonically tailored dipole potentials and nanofibers \cite{nanofibers1,nanofibers2,nanofibers3,nanofibers4} offer promising tools for trapping, guiding and manipulating atoms. In the present publication we investigate also under what conditions the artificial magnetic fields 
created by 
evanescent waves can be responsible for reflection of atoms falling down on a surface of a prism at temperatures higher than the critical value for the quantum degeneracy.


\section*{Results}

\noindent
{\bf Adiabatic motion of atoms.}
We consider a two-level atom interacting with an external laser field. Assume that the atomic energy level difference is $\hbar\omega_0$ and the atom at rest is located at $\vect{r}$. The Hamiltonian of the internal degrees of freedom of the atom interacting with an electromagnetic field is time periodic due to the periodicity of the electromagnetic wave, i.e. $H_{in}(t+2\pi/\omega)=H_{in}(t)$. According to the Floquet theorem \cite{floquet883,shirley65,zeldovich66} (an analogue of the Bloch theorem in solid-state physics), the operator (the so-called Floquet Hamiltonian) $\mathcal{H}_{in}=H_{in}-i\hbar\partial_t$ possesses time periodic eigenstates. The corresponding eigenvalues are defined modulo $\hbar\omega$ and are called quasi-energies. Just as in the solid-state physics, one can reduce considerations to a single Floquet zone (the equivalent of the Brillouin zone). Eigenstates $|\chi_1(\vect{r})\rangle$ and $|\chi_2(\vect{r})\rangle$ of the Floquet Hamiltonian for the two-level atom problem, i.e. the 
so-called dressed states in the atomic optics context, can be found analytically if the rotating wave approximation (RWA) is applied. The quasi-energy splitting equals  $\epsilon_1(\vect{r})-\epsilon_2(\vect{r})=\hbar\Omega(\vect{r})$ where $\Omega(\vect{r})=\sqrt{\Delta^2+|\kappa(\vect{r})|^2}$ is the generalized Rabi frequency. $\Delta=\omega_0-\omega$ is the detuning from the resonance frequency and $\kappa(\vect{r})={\bf d}\cdot{\bf E}(\vect{r})/\hbar$ is the Rabi frequency where $\vect{d}$ and $\vect{E}(\vect{r})$ stand for the atomic dipole moment and the electric field vector, respectively. In the present paper we consider small detuning and neglect spontaneous emission of an atom. Therefore, the presented results are relevant to, e.g., the long-lived clock transition in ytterbium atoms \cite{hinkley2013}.

If we assume that the atom is initially prepared in, e.g., the $|\chi_1(\vect{r})\rangle$ dressed state and the atom moves sufficiently slowly, the dressed state is adiabatically followed by the atom. Then, non-trivial vector ${\bf A}$ and scalar $W$ potentials can appear in the Hamiltonian that describes the motion of the center of mass of the atom and the system can mimic dynamics of a charged particle in the presence of a magnetic field \cite{juzel2008,dalibard2011,juzeliunas13}. The vector potential ${\bf A}$ is the consequence of the Berry phase \cite{berry1984} that emerges due to the adiabatic approximation, i.e.
\be 
\gamma(C)=i\oint_C\langle\chi_1|\nabla\chi_1\rangle\cdot d{\bf r}=\frac{1}{\hbar}\oint_C{\bf A}\cdot d{\bf r}.
\label{berry}
\ee
Thus, the vector potential $\vect{A}$ can be expressed in the form
\be
\textbf{A}=i\hbar\langle\chi_1|\nabla\chi_1\rangle,
\label{vecpot}
\ee
and the corresponding magnetic field, ${\vect{B}}=\nabla\times {\vect{A}}$, depends on the gradient of the phase of the external electromagnetic wave and the gradient of the generalized Rabi frequency $\Omega(\vect{r})$. The scalar potential $W$ reads
\be
W=\frac{\hbar^2}{2m}|\langle\chi_2|\nabla\chi_1\rangle|^2.
\label{skalarpot}
\ee
The artificial gauge potentials are called 'geometrical' because they depend only on the spatial variation of the dressed state as one can see in (\ref{vecpot}) and (\ref{skalarpot}).

In the following we consider a total internal reflection of light in a prism that creates an evanescent wave. An evanescent wave seems to have all important properties necessary to generate artificial gauge potentials for adiabatically moving atoms, i.e. it has a gradient of the phase and a large gradient of the amplitude. 

\begin{figure}[h]
\centering
\includegraphics*[scale=3.5]{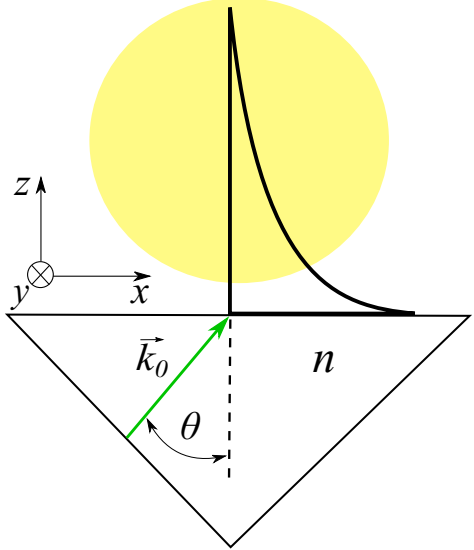}
\caption{(Color online) The geometry of the considered system. A plane wave with the wave vector $\vect{k}_0$ (green arrow) strikes the prism surface at the incident angle $\theta$ greater than the critical angle for the total internal reflection. The resulting evanescent wave interacts with a cloud of atoms (yellow circle) trapped near the prism surface.}
\label{system}
\end{figure}

\noindent
{\bf Bose-Einstein condensate in an artificial magnetic field.}
Let us consider a prism made of dielectric material with the refractive index $n>1$ and an electromagnetic plane wave that propagates in the prism. The wave strikes the boundary between the dielectric medium and the vacuum at the angle of incidence $\theta$ greater than the critical angle $\theta_0=\textrm{arcsin}(1/n)$ for the total internal reflection. The evanescent wave, that appears in the vacuum, propagates along the boundary ($x$ direction) and decays exponentially with increasing distance from the boundary ($z$ direction), see Fig.~\ref{system},
\begin{equation}
	{\vect{E}}(x,z,t)=t^{TE}(\theta)\;\vect{E}_0\;e^{-i\omega t}\;e^{i\phi(x)}\;e^{-z/d},
\label{singlee}
\end{equation}
where $\vect{E}_0$ describes the amplitude and the direction of the electric field vector, $\phi(x)=xk_0n\sin\theta$ is the running phase, $d=\left(k_0\sqrt{n^2\sin^2\theta-1}\right)^{-1}$ is the penetration depth and $k_0=2\pi/\lambda$ is the wavenumber. We have chosen the $TE$ polarization but similar analysis can be performed for the $TM$ polarization. The transmission coefficient is $t^{TE}(\theta)=2n\cos\theta\left(n\cos\theta+i\sqrt{n^2\textrm{sin}^2\theta-1}\right)^{-1}$.

The dressed states of a two-level atom in the presence of such an evanescent field, obtained within the RWA, read 
\be
	|\chi_1(x,z)\rangle=\left( {\begin{array}{cc}
\cos[\Phi(z)/2]   \\
\sin[\Phi(z)/2]\;e^{-i\phi(x)}  \\
\end{array} } \right),
\ee
\be
	|\chi_2(x,z)\rangle=\left( {\begin{array}{cc}
-\sin[\Phi(z)/2]\;e^{i\phi(x)} \\
\cos[\Phi(z)/2] \\
\end{array} } \right),
\ee
with energies $\hbar\Omega(z)/2$, $-\hbar\Omega(z)/2$ respectively and $\Phi(z)=\textrm{arctg}(|\kappa(x,z)|/\Delta)$.
We assume that slowly moving atoms follow one of the dressed states, e.g. $|\chi_1(x,z)\rangle$. It is possible because the energy split in the dressed atom picture is $\hbar\Omega$ which leads to the separation of the dynamics of each dressed state and allows for the adiabatic elimination of one of them. The condition for the applicability of the adiabatic approximation can be obtained by rewriting the internal state $\Psi({\bf r}(t))=\sum_i{c_i(t)|\chi_i({\bf r}(t))\rangle}$ and solving the Schr\"odinger equation as a power series in velocity \cite{cheneau}. Then the adiabatic motion requires $|c_2|\ll1$ that gives the range of velocities $v\ll \Omega/|\langle\chi_2|\bigtriangledown\chi_1\rangle|$. The vector potential associated with the adiabatic motion is then
\be
{\vect{A}}(x,z)=\hbar\sin^2\left[\Phi(z)/2\right]\nabla\phi(x).
\ee
The calculation of the curl of the vector potential $\vect{A}$ allows one to obtain the artificial magnetic field vector, which has a nonzero component in the $y$ direction only,
\be
\vect{B}(z)=-\hat y B(z)=-\hat{y}B_0 \sqrt{n^2\sin^2\theta-1}\frac{s^2 \alpha(z)n\sin\theta }{\left[1+s^2\alpha(z)\right]^{3/2}},
\label{bplanew}
\ee
where $B_0=\hbar k^2_0/2$ and $\alpha(z)=\left|t^{TE}(\theta)\right|^2 e^{-2z/d}$, and we have introduced the parameter \be
s=\frac{|\vect{d}\cdot\vect{E}_0|}{\hbar|\Delta|}.
\label{s}
\ee

\begin{figure}[h]
\centering
\includegraphics*[scale=0.8]{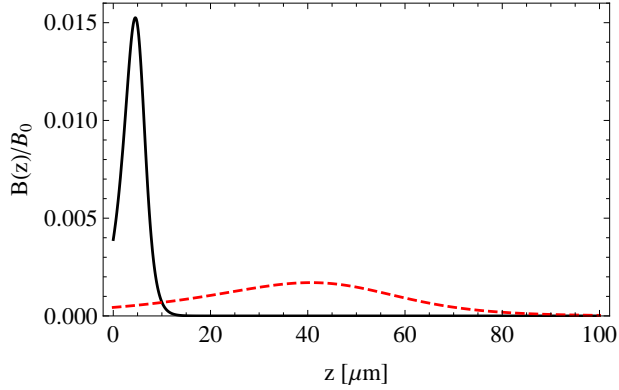}
\caption{(Color online) The magnetic field $B(z)$ created by a plane wave, in units of $B_0=\hbar k_0^2/2$, as a function of $z$ for two different angles of the incidence $\theta-\theta_0=8\cdot10^{-4}$rad (solid black line), $\theta-\theta_0=10^{-5}$rad (dashed red line) and for $s=5$, see Eq.~(\ref{s}), and $\lambda=578$~nm.}
\label{ev1}
\end{figure}
\begin{figure}[h]
\centering
\includegraphics*[scale=0.799]{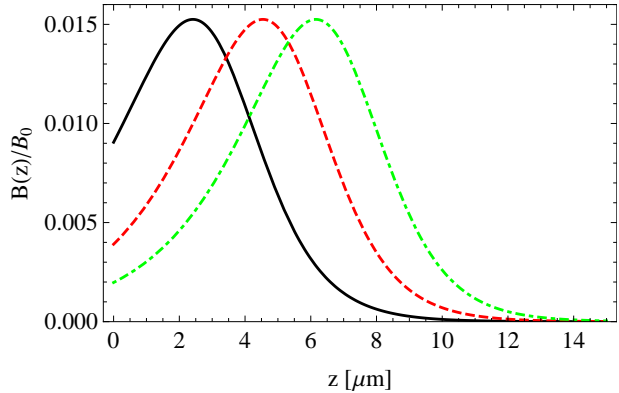}
\caption{(Color online) The magnetic field $B(z)$ created by a plane wave, in units of $B_0=\hbar k_0^2/2$, as a function of $z$ for three different values of the parameter $s$, Eq.~(\ref{s}), i.e. $s=2$ (solid black line), $s=5$ (dashed red line) and $s=10$ (green dotted-dashed line), and for $\theta-\theta_0=8\cdot10^{-4}$rad and $\lambda=578$~nm.
}
\label{ev2}
\end{figure}
\begin{figure}[h]
\centering
\includegraphics*[scale=0.8]{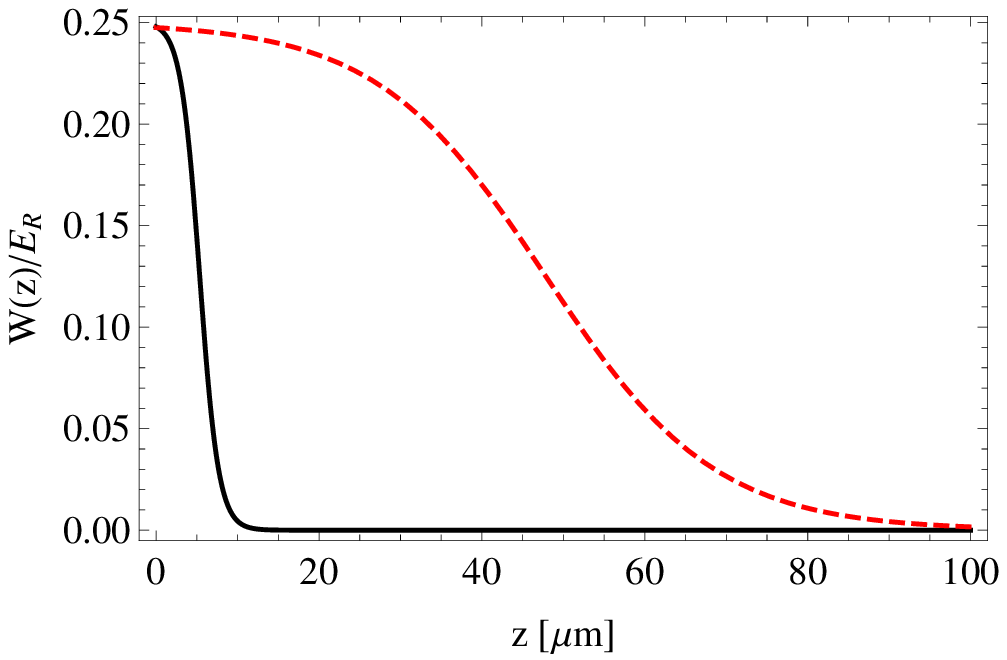}
\caption{(Color online) The geometrical scalar potential $W(z)$ created by a plane wave, in units of the energy recoil $E_R=\hbar^2k_0^2/(2m)$, as a function of $z$ for two different angles of incidence $\theta-\theta_0=8\cdot10^{-4}$rad (solid black line), $\theta-\theta_0=10^{-5}$rad (dashed red line) and for $s=5$, see Eq.~(\ref{s}), and $\lambda=578$~nm.}
\label{Ugeom}
\end{figure}

The magnetic field $B(z)$ can be shaped by changing $\theta$. The angle of incidence determines both the maximal value of the magnetic field and a range $\Delta z$ over which $B(z)$ is significant. The strongest field is created if the incident angle of the plane wave $\theta$ is much greater than the critical angle for the total internal reflection $\theta_0$. Then, however, the magnetic field is present on a small region $\Delta z\approx d\approx 1/k_0$. Integrating the artificial magnetic field, Eq.~(\ref{bplanew}), we obtain
\be
\int\limits_0^{+\infty}B(z)dz=\hbar k_0\sin^2\left[\Phi(0)/2\right]\le \hbar k_0.
\label{bint}
\ee
Thus, the maximal magnetic field is of the order of $\hbar k_0^2$. When $\theta$ approaches $\theta_0$, the penetration depth $d$ increases but the artificial magnetic field becomes weaker $B\propto 1/d$. 

In order to decide which values of $\theta$ are suitable for an experiment one should analyze how many vortices in ultra-cold atomic gases the artificial magnetic field is able to create. The vortex density can be expressed by $\rho_v=B/(2\pi\hbar)$. If $B$ keeps significant value in a square of area $(\Delta z)^2$, the number of vortices in this square is $(\Delta z)^2\rho_v$. The field $B(z)$ depends on $z$ coordinate only, thus, the space where the magnetic field is significant forms a layer of width $\Delta z$. Therefore, it is more instructive to estimate the number of vortex rows $N_{rows}=\Delta z \sqrt{\rho_v}$ which, for $\theta$ close to the critical angle $\theta_0$, can be approximated by
\be
N_{rows}\approx\frac{1}{2}\sqrt{\frac{d}{\lambda}}\approx\left(\frac{1}{8\sqrt{2}\pi(n^2-1)^{1/4}\sqrt{\theta-\theta_0}}\right)^{1/2}.
\label{Nrows}
\ee
For $n=1.4$ and $\lambda=578$~nm we obtain $N_{rows}=1$ and $\Delta z\approx 2.3$~$\mu$m if $\theta-\theta_0\approx 8\cdot 10^{-4}$rad while if $\theta-\theta_0\approx 10^{-5}$rad, $N_{rows}=3$ and $\Delta z\approx 20.8$~$\mu$m. The adjustment of the incidence angle with accuracy of $10^{-4}$rad has been demonstrated in a laboratory \cite{cornelussen2002} but much better accuracy can be attainable in evanescent wave experiments.

In Fig.~\ref{ev1} we show the plots of the magnetic fields $B(z)$ that correspond to these two examples. By changing $\theta$ one can control how many vortex rows are realized experimentally and then investigate how the spatial arrangement of vortices changes with an increase of $N_{rows}$. Atoms are charge neutral but to get a sense of the order of magnitude of the artificial magnetic field created by means of the evanescent wave let us assume that atoms possess the elementary charge $e$. Then, the black curve in Fig.~\ref{ev1} corresponds to magnetic field $B/e\approx 0.3$mT that is present on the region $\Delta z\approx 10\mu$m.

An increase of the parameter $s$, Eq.~(\ref{s}), causes essentially a shift of $B(z)$ towards greater values of $z$, see Fig.~\ref{ev2}. Thus, the location of the region where the artificial magnetic field is present can be suitably chosen by a change of the parameter $s$. It allows one to trap atomic clouds sufficiently far away from the surface of the prism and consequently eliminate the influence of the van der Waals interaction between the atom and the dielectric wall \cite{westbrook98}. 

In order to trap atoms close to the surface of the prism an external magnetic trap or additional laser beams have to be applied \cite{cornelussen2002}. The geometric potential $W$ is too weak to overcome the gravitational attraction. In Fig.~\ref{Ugeom} we show 
\bea
W(z)&=&\frac{\hbar^2}{8m}\left(\frac{1}{d^2}\frac{s^2\alpha(z)}{[1+s^2\alpha(z)]^2}+\frac{s^2\alpha(z)}{1+s^2\alpha(z)}k_0^2n^2\textrm{sin}^2\theta\right), \cr &&
\eea
for parameters corresponding to those used in Fig.~\ref{ev1}. For example for ytterbium atoms, the maximal force created by this potential is $0.17mg$ where $g$ is the gravitational acceleration. Also the optical dipole potential created by the considered evanescent waves can be too weak to keep an atomic cloud above the surface of a prism. Indeed, the artificial magnetic fields suitable for experiments require large penetration depth $d$. However, by increasing $d$ we decrease the optical dipole force because $\nabla \Omega$ becomes smaller.


If the incidence angle of a laser beam is significantly greater than the critical angle $\theta_0$, it is possible to approximate the beam by a plane wave. The situation changes for the incident angles close to $\theta_0$ and such a situation is investigated in the present publication. Therefore we have to consider the realistic laser radiation in the analysis of the artificial magnetic fields induced by evanescent waves.

We consider a laser beam incident with an angle $\theta_{in}$ on a boundary between a dielectric medium and the vacuum. The beam is represented by a Gaussian superposition of plane waves. The resulting electric field in the vacuum is a sum of two contributions
\be
\vect{E}(\vect{r},t)=\vect{E}_1(\vect{r},t)+\vect{E}_2(\vect{r},t).
\label{sumE}
\ee
The first contribution describes the evanescent field, i.e. it corresponds to the superposition of plane waves incident with angles $\theta>\theta_0$, 
\bea
\vect{E}_1(\vect{r},t)&=&\frac{\vect{E}_0e^{-i\omega t}}{\sqrt{\pi}\Delta\theta}\int^{\pi/2}_{\theta_0}d\theta\;t^{TE}(\theta)e^{i\phi(x)}e^{-z/d} \cr &\times& \exp\left[ink_0\frac{l}{2}(\theta-\theta_{in})^2-\frac{(\theta-\theta_{in})^2}{(\Delta\theta)^2}-\frac{y^2}{w_y^2}\right], \cr &&
\label{e1e}
\eea
where the last exponential term describes the profile of the beam. $l$ is the distance of the waist of the incident beam from the surface, $\Delta\theta=2/(nk_0w)$ describes the Gaussian distribution of the incident angles, where $w$ is the waist of the beam, and $w_y$ is the radius of the transverse distribution \cite{cornelussen2002}. The second contribution in Eq.~(\ref{sumE}), i.e. $\vect{E}_2(\vect{r},t)$, describes the propagation of waves that strike the surface with $\theta<\theta_0$ and it is given by the similar formula as Eq.~(\ref{e1e}) but the range of the integration is between 0 and $\theta_0$ and $\exp\left(izk_0\sqrt{1-n^2\sin^2\theta}\right)$ substitutes for $e^{-z/d}$.

\begin{figure}[h]
\centering
\includegraphics*[scale=0.8]{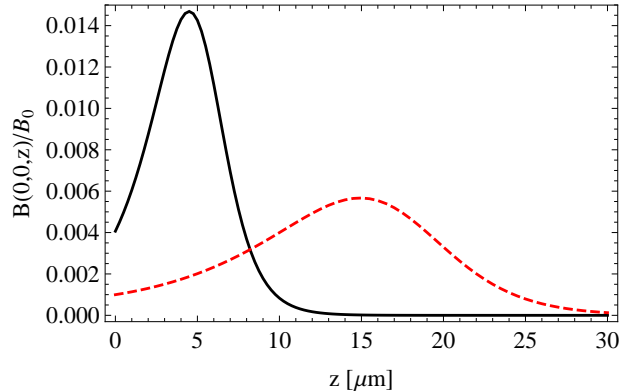}
\caption{(Color online) Magnetic field $B(0,0,z)$ created by a Gaussian laser beam, in units of $B_0=\hbar k_0^2/2$, as a function of $z$ for two different angles of incidence $\theta_{in}-\theta_0=8\cdot10^{-4}$rad (solid black line), $\theta_{in}-\theta_0=10^{-5}$rad (dashed red line) and for $s=5$, see Eq.~(\ref{s}). The parameters of the laser beam are as follows: $\lambda=578$~nm, $l=680$~mm, $w=440\;\mu$m and $w_y=440\;\mu$m, see Eq.~(\ref{e1e}).}
\label{Bbeam}
\end{figure}

Let us assume that the laser beam has experimentally realistic parameters: $\lambda=578$~nm, $l=680$~mm, $w=440\;\mu$m and $w_y=440\;\mu$m \cite{cornelussen2002}. We consider the incident angles for which $d(\theta_{in})\ll w_y$ and consequently the Rabi frequency changes much more slowly with the change of $y$ than $z$. Thus, the dominant component of the resulting artificial magnetic field vector is the $y$-component, i.e. $\vect{B}(\vect{r})\approx -B(\vect{r})\hat y$ similarly as in the previous section. A new feature for the present configuration is that the shape of $B(\vect{r})$ depends both on the incident angle $\theta_{in}$ and on the parameter $s$.  

In Fig.~\ref{Bbeam} we show $B(0,0,z)$ versus $z$ for $s=5$ and for the same incident angles as in the case of a single plane wave, cf. Fig.~\ref{ev1}. For $\theta_{in}-\theta_0=8\cdot10^{-4}$rad the artificial magnetic field is nearly the same as in Fig.~\ref{ev1} and consequently the plane wave approximation of a laser beam is valid. For $\theta_{in}-\theta_0=10^{-5}$rad there is a noticeable difference between the cases of the single plane wave and the realistic laser beam. That is, the maximal field is greater but the range of $\Delta z$ where the magnetic field is significant is smaller. 


So far we have analyzed the creation of the artificial magnetic field by means of an evanescent wave and estimated number of vortices that can be formed in ultra-cold atomic gases in the presence of such fields. In order to confirm the predictions we switch to the numerical simulations within the mean field approximation. We consider a Bose-Einstein condensate trapped in a harmonic potential in the presence of the vector potential $\vect{A}$ in the two-dimensional (2D) approximation. For the parameters we are going to use, the geometrical scalar potential (\ref{skalarpot}) and the optical dipole potential are very weak and therefore are neglected. \ The Gross-Pitaevskii equation (GPE) in the units of the harmonic trapping potential reads
\bea
\mu\psi &=&-\frac12\left[\left(\partial_x+iA_x\right)^2+\left(\partial_z+iA_z\right)^2\right]\psi \cr &&
+\frac{x^2+z^2}{2}\psi+g|\psi|^2\psi, 
\label{gpe}
\eea
where $\mu$ is the chemical potential of the system and $g$~stands for the atomic interaction strength (we assume $\langle\psi|\psi\rangle=1$). In the numerical simulation we discretize 2D space. A naive approximation of, e.g., $\partial_x\psi(x,z)\approx [\psi(x+dx,z)-\psi(x-dx,z)]/(2dx)$ leads to a discrete version of the GPE which is not gauge invariant. This problem can be overcome by~adopting the Schwinger line integral used in~the lattice gauge theories \cite{lattice_gauge}. That is, in order to make a discrete version of the energy functional gauge invariant, terms like $\psi^*(x,z)\psi(x+dx,z)$ have to be exchanged by gauge invariant terms
\be
\psi^*(x,z)\;U(x,z;x+dx,z)\;\psi(x+dx,z),
\ee
where the Schwinger line integral $U(x,z;x+dx,z)=\exp(iA_xdx)$. It corresponds to the following substitution in the GPE
\bea
\left(\partial_x+iA_x\right)^2\psi(x,z)&\longrightarrow& \frac{1}{dx^2}[U\psi(x+dx,z)\cr &&+ U^*\psi(x-dx,z)-2\psi(x,z)], \cr &&
\eea
and similarly for $\left(\partial_z+iA_z\right)^2\psi$.
The resulting discrete GPE is gauge invariant and recovers Eq.~(\ref{gpe}) in the limit where $dx$ and $dz$ go to zero. 

\begin{figure}[h]
\centering
\includegraphics*[scale=0.32]{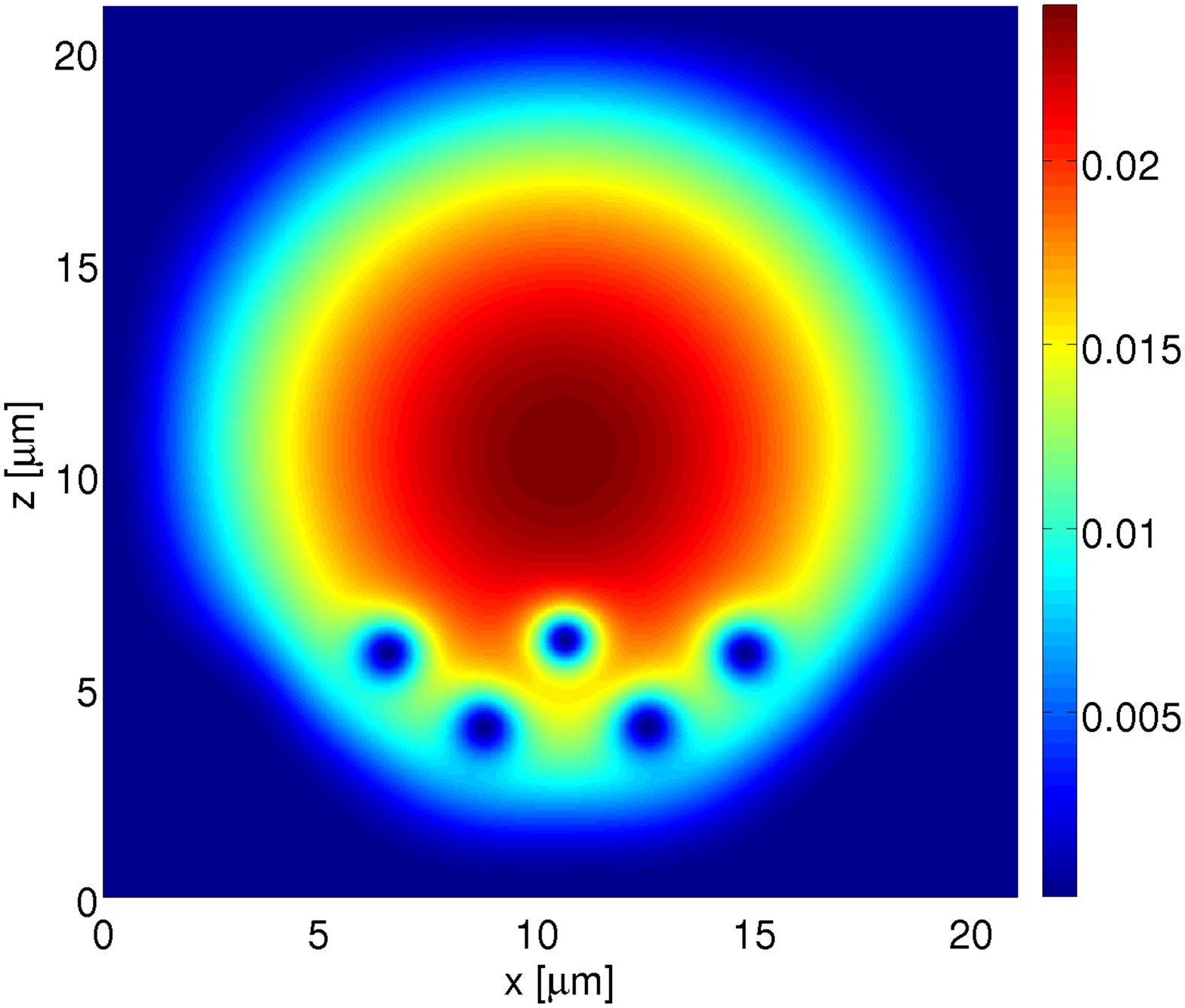}
\includegraphics*[scale=0.32]{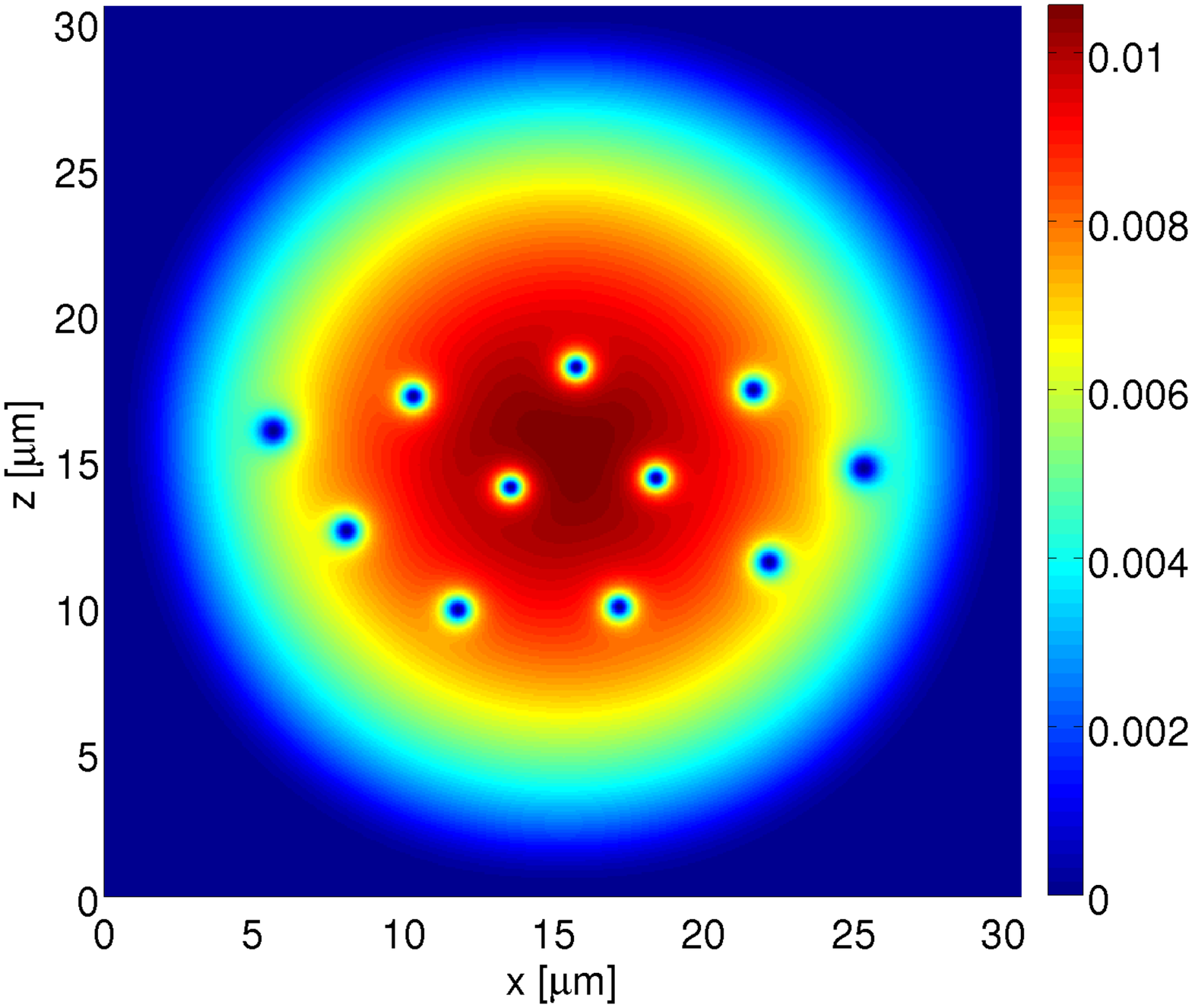}
\caption{(Color online) The probability density of ytterbium atoms trapped in the harmonic potential corresponding to the frequency $\omega_{trap}=2\pi\times 16$Hz. The interaction coefficient $g$, see Eq.~(\ref{gpe}), has been chosen so that the Thomas-Fermi radius of a cloud of ytterbium atoms is $R_{TF}=10\mu$m (top panel) and $R_{TF}=15\mu$m (bottom panel). The 2D space has been discretized, i.e. $dx=dz=0.05\mu$m (top panel) and $0.063\mu$m (bottom panel). Vortices visible in the plots are the results of the presence of the artificial magnetic field created by the evanscent wave. The parameters of the magnetic field are the same as in Fig.~\ref{Bbeam}, i.e. the top panel of the present figure is related to the black curve in Fig.~\ref{Bbeam} while the bottom panel to the red curve. }
\label{num}
\end{figure}

The ground states of the GPE found numerically for the artificial magnetic field created by means of the evanescent wave with the Gaussian profile are presented in Fig.~\ref{num}. Top and bottom panel in Fig.~\ref{num} corresponds to the black and the red curve in Fig.~\ref{Bbeam}, respectively. The interaction coefficient $g$ has been chosen so that the Thomas-Fermi radius of a cloud of ytterbium atoms, in the harmonic potential with the frequency $\omega_{trap}=2\pi\times 16$Hz, is $R_{TF}=10\mu$m (top panel) and $R_{TF}=15\mu$m (bottom panel). Many different combinations of the total particle number and the trap frequency along the third direction lead to the same coefficient $g$ in the 2D approximation. Therefore, it is meaningful to provide $g$, or $R_{TF}$ equivalently, only and do not consider a particular choice of the particle number and the trap frequency in the $y$ direction. 

We analyze the behaviour of the system in the Thomas-Fermi regime. Thus, the healing length of the system, and consequently the size of the vortex cores, is much smaller than $R_{TF}$ and also smaller than the region of space $\Delta z$ where the artificial magnetic field is significant. The numbers of vortex rows that can be estimated according to
\be
N_{rows}\approx \Delta z\sqrt{\frac{B}{2\pi\hbar}},
\ee
where $B$ is half of the maximum value of the magnetic field are 1 and 2, respectively, which is smaller than the actual numbers visible in Fig.~\ref{num}. It means that our estimate is even too pessimistic. The reason is that in the estimation we take the minimal value of the magnetic field in the considered region but in fact atoms feel stronger field.


\noindent
{\bf Cold atoms in an artificial magnetic field.}
We would like to analyze now if the presence of artificial magnetic fields induced by evanescent waves can be observed in experiments with cold atomic gases at temperatures higher than the critical value for the quantum degeneracy. We will deal with velocities of atoms of the order of a meter per second which are much greater than in the case of ultra-cold gases. To fulfill the adiabaticity condition it is necessary to apply a sufficiently large detuning $\Delta$. This in turn implies a large Rabi frequency because in order to have a significant artificial magnetic field the parameter $s$ in Eq.~(\ref{s}) must be at least of the order of unity. Such conditions can be difficult to fulfill for ytterbium atoms. Therefore, in the present section we consider a different arrangement for creation of artificial magnetic fields. 

Let us consider two degenerate internal states $|1\rangle$ and $|2\rangle$ that belong to the ground-state manifold of, e.g., $^{87}$Rb atoms and two laser beams, characterized by the Rabi frequencies $\kappa_1(x,z)$ and $\kappa_2(x)$ and the wave vectors $k_1\approx k_2\equiv k_0$,  which can couple these states to an excited state. We assume large detuning of the laser radiations from the resonant transition and $\gamma=(|\kappa_1|^2+|\kappa_2|^2)/\Delta^2\ll 1$. The first beam strikes a surface of a prism with an incident angle $\theta$ greater but very close to the critical angle $\theta_0$ for the total internal reflection and produces an evanescent wave, see Fig.~\ref{system}. The other beam propagates in the vacuum along the prism surface, i.e. with the wave vector $\vect{k}_2=-k_0\hat x$. The laser beams can induce Raman transitions between the two ground states. We assume that atoms follow adiabatically the dressed state 
\be
|\chi_1\rangle=\frac{|1\ra+\zeta|2\rangle}{\sqrt{1+|\zeta|^2}}, 
\label{darks}
\ee
where $\zeta=-\kappa_1^*/\kappa_2^*{=-\tilde{s}e^{-z/d}e^{-ik_0(n\sin\theta+1)x}}$ and $\tilde s=\left|\vect{d}_1\cdot\vect{E}_{{0}1}\right|/\left|\vect{d}_2\cdot\vect{E}_{{0}2}\right|$ \cite{fleischhauer2005}. $\vect{E}_{0i}$ describe amplitudes and directions of the electric field vectors of the laser beams and $\vect{d}_{i}$ stand for the atomic dipole moments. Then, in the Hamiltonian that describes the center of mass motion of an atom, geometric vector, $\vect{A}=i\hbar\langle\chi_1|\nabla\chi_1\rangle$, and scalar, $W=\hbar^2|\langle\chi_2|\nabla\chi_1\rangle|^2/(2m)$, potentials are present \cite{dalibard2011}, where  $|\chi_2\rangle$ is the nearest neighbour dressed state separated in energy by $\hbar\delta\epsilon=-\hbar\Delta\gamma/4$. For a given value of $\gamma$, the detuning $\Delta$ can be chosen so large that the adiabaticity condition $|v|\ll|\delta\epsilon|/|\langle\chi_2|\bigtriangledown\chi_1\rangle|$ is fulfilled even if atoms move with velocities $v$ of the order of a meter per second. 
The resulting artificial magnetic field reads
\be
\vect{B}(z) = 
-\hat{y}\hspace{0.1em}2\hbar k_0^2(n\sin\theta+1)\sqrt{n^2\sin^2\theta-1}\frac{\tilde{s}^2e^{-2z/d}}{(1+\tilde{s}^2e^{-2z/d})^2},
\ee
and it is significant if the parameter ${\tilde s}$ 
is of order of unity or greater. 

\begin{figure}[h]
\centering
\includegraphics*[scale=0.32]{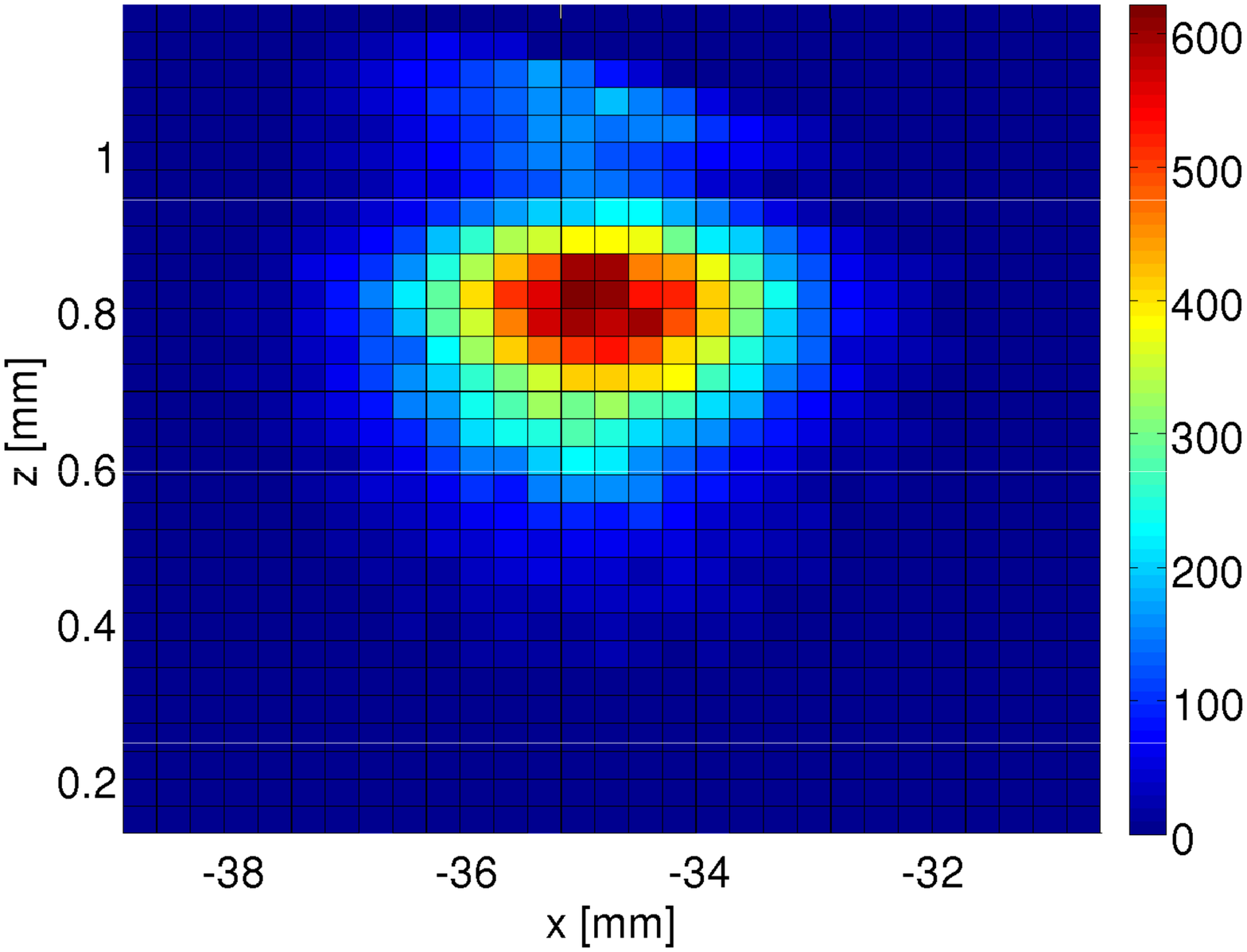}
\caption{ (Color online) Density of atoms after the release form the magneto-optical trap and $35$~ms evolution in the presence of the gravitational field and the artificial magnetic field induced by an evanescent wave. The surface of the prism is oriented horizontally along $x$ axis and located at $z=0$, cf. Fig.~\ref{system}. The gravitational force is oriented along $z$ axis and points downwards. The results are obtained in the classical trajectory simulations. Initially $N = 10^6$ $^{87}$Rb atoms, at $T=10\;\mu$K, are prepared in a magneto-optical trap ($\omega_{trap} = 2\pi\times100$~Hz) at $z=1$~mm above the surface of the prism and at $x=0$. It is assumed that after the trap is turned off, atoms are accelerated to the average velocity $\vect{v} = -1\hat x$~m/s. Next, the cloud of atoms moves in the gravitational field and falls down on the dielectric surface. Atoms feel an artificial magnetic field induced by two red-detuned ($\Delta = 10$~GHz, $\lambda = 795.5$~nm) laser beams with the ratio of the 
Rabi frequency $\tilde s=5$. The first beam propagates in the prism and strikes its surface at the incident angle $\theta = \theta_0+3\cdot 10^{-4}$~rad and creates an evanescent wave. The other one propagates in the vacuum along the $x$ direction. We assume that atoms are prepared in the dressed state (\ref{darks}) which uncouples from an atomic excited state. The only scalar potentials experienced by atoms correspond to the geometrical potential $W$ and the gravitational one. The former creates too weak force (smaller than $0.2mg$) to overcome the gravitational field and is not able to reflect atoms. Note that in the absence of the artificial magnetic field all atoms are lost at the prism surface after evolution time shorter than 35 ms. If the artificial field is on, 5\% of atoms are reflected from the prism surface. 
} 
\label{mot}
\end{figure}

We consider a cloud of atoms prepared in a magneto-optical trap 1~mm above the horizontally oriented prism surface (cf. Fig.~\ref{system}), at temperature $T=10\;\mu$K --- in Fig.~\ref{system}, the gravitational force is oriented along $z$ axis and points downwards. We assume that after the release of atoms from the trap they are accelerated to the average velocity $\vect{v}=-1 \hat{x}$ m/s. With such initial conditions the cloud of atoms expands in the gravitational field and falls down on the surface of the prism. Artificial magnetic field is created by an evanescent wave which penetrates space above the prism.
The dressed state (\ref{darks}) uncouples from an atomic excited state and there is no light shift of this level. The geometrical scalar potential $W$ is too weak to reflect the falling atoms. Thus, an atom does not hit the dielectric surface only if the artificial magnetic field is able to bend its trajectory due to the artificial Lorentz force. Figure~\ref{mot} shows density of atoms 35~ms after the release from the trap obtained in classical trajectory simulations. If the artificial magnetic field was not present, such a time would be sufficient for all atoms to fall down and strike the dielectric surface. Due to the dispersion of initial atomic velocities (i.e. $\Delta v\approx0.03$~m/s for $T=10\;\mu$K), many atoms hit the prism surface and are lost. However, some fraction of atoms (i.e. those which enter the region of the magnetic field with a sufficiently small angle with respect to the region boundary) are able to bounce repeatedly on the prism surface due to the presence of the synthetic magnetic 
field. Figure~\ref{mot} shows the atomic density at the moment of time corresponding to the turning point of atoms. At $t=$35~ms atoms reach the highest positions, return, bounce off the dielectric surface and again reach the highest position at $t\approx$70~ms and so on.

In the example illustrated in Fig.~\ref{mot} we have assumed that after the atomic trap is turned off, atoms are accelerated to the average velocity $\vect{v} = -1\hat x$~m/s. It is necessary in order to ensure that atoms will enter the region of the artificial magnetic field with small angles with respect to the region boundary. Then, the Lorentz force is able to bend atomic trajectories and reflect atoms from the prism surface. 
Such an additional acceleration can be omitted in an experiment if the surface of the prism is not oriented horizontally but forms angle of about 0.1~rad with the gravitational force vector. Then, the initial acceleration is not necessary because the gravitational field accelerates atoms to the suitable velocities.

\section*{Discussion}
\label{conclusions}

We have considered atoms that move slowly in the presence of an evanescent wave. The theoretical description of the adiabatic atomic motion involves the geometrical Berry phases that can be represented by vector and scalar potentials experienced by atoms. Such artificial gauge potentials are the stronger the greater gradients of the phase and amplitude of an external electromagnetic field \cite{dalibard2011}. An evanescent wave is a good candidate for the realization of the gauge fields due to an exponential decay of its amplitude. 

We analyze two laser beam configurations that lead to synthetic magnetic fields. The first configuration assumes creation of an evanescent wave by means of laser radiation which is nearly resonant with an electronic transition. This method can be applied to atoms with a large radiative lifetime. If a Bose-Einstein condensate is placed in such an evanescent wave, the synthetic magnetic field can induce vortices in the condensate. In order to create a large number of vortices the angle of incidence has to be very close to the critical angle for the total internal reflection. Then, realistic profile of a laser beam has to be taken into account. We show that an evanescent wave with the Gaussian profile with experimentally attainable parameters is able to create an unidirectional magnetic field. 

We also analyze artificial magnetic fields induced by two laser beams. The first laser radiation creates an evanescent wave while the other beam propagates in the vacuum and together with the first one can induce Raman transitions. Atoms prepared in a dressed state which uncouples from an excited state experience a synthetic magnetic field and are not affected by a radiative decay. We illustrate this method with a system of a cold atomic gas at temperature higher than the critical value for the quantum degeneracy. The presence of the artificial magnetic field influences motion of an atomic cloud that is released from a magneto-optical trap and falls down on a dielectric surface. In such a way one can realize a new type of the atomic mirror that is based on the artificial Lorentz force and this effect can be measured experimentally.

After the present article appeared in the arXiv basis, an article on a similar topic was submitted to arXiv and recently published \cite{lembessis}.



\section*{Acknowledgments}
We are grateful to Tomasz Kawalec for fruitful discussions and suggestions related to cold atoms experiments. 
This work is supported by National Science Centre via project numbers DEC-2012/05/N/ST2/02745 (MM) and DEC-2012/04/A/ST2/00088 (KS).

\section*{Author Contributions}
The idea of the research was established in the joined discussion of MM and KS. MM performed all calculations under the supervision of KS. Both authors contributed in the article preparation.

\section*{Additional Information}
{\bf Competing financial interests:} The authors declare no competing financial interests.

\end{document}